\documentclass[fleqn,10pt,twocolumn]{wlscirep}
\usepackage[utf8]{inputenc}
\usepackage[T1]{fontenc}
\usepackage{multirow}
\usepackage{tabularx}
\usepackage{tabularray}
\usepackage{color}
\usepackage{algorithm}
\usepackage{algpseudocode}
\title{Creating an Atlas of Normal Tissue for Pruning WSI Patching Through Anomaly Detection}
\author[1]{Peyman Nejat}
\author[1]{Areej Alsaafin}
\author[1]{Ghazal Alabtah}
\author[2]{Nneka Comfere}
\author[3]{Aaron Mangold}
\author[1, 2]{Dennis Murphree}
\author[4]{Patricija Zot}
\author[4]{Saba Yasir}
\author[4]{Joaquin J. Garcia}
\author[1]{H.R. Tizhoosh}
\affil[1]{Rhazes Lab, Department of Artificial Intelligence and AI, Mayo Clinic, Rochester, Minnesota, 55901, USA}
\affil[2]{Department of Dermatology, Mayo Clinic, Rochester, MN, USA}
\affil[3]{Department of Dermatology, Mayo Clinic, Scottsdale, AZ, USA}
\affil[4]{Department of Laboratory Medicine and Pathology, Mayo Clinic, Rochester, MN, USA}
\begin{abstract}
Patching gigapixel whole slide images (WSIs) is an important task in computational pathology. Some methods have been proposed to select a subset of patches as WSI representation for downstream tasks. While most of the computational pathology tasks are designed to classify or detect the presence of pathological lesions in each WSI, the confounding role and redundant nature of normal histology in tissue samples are generally overlooked in WSI representations. In this paper, we propose and validate the concept of an ``atlas of normal tissue'' solely using samples of WSIs obtained from normal tissue biopsies. Such atlases can be employed to eliminate normal fragments of tissue samples and hence increase the representativeness collection of patches. We tested our proposed method by establishing a normal atlas using 107 normal skin WSIs and demonstrated how established indexes and search engines like Yottixel can be improved. We used 553 WSIs of cutaneous squamous cell carcinoma (cSCC) to show the advantage. We also validated our method applied to an external dataset of 451 breast WSIs. The number of selected WSI patches was reduced by 30\% to 50\% after utilizing the proposed normal atlas while maintaining the same indexing and search performance in leave-one-patinet-out validation for both datasets. We show that the proposed normal atlas shows promise for unsupervised selection of the most representative patches of the abnormal/malignant WSI lesions.
\end{abstract}
\begin{document} 
\flushbottom
\maketitle
\thispagestyle{empty}
\section*{Introduction}
Pathology is the field of studying the cause, development, and effects of diseases. This field involves examining sampled tissues, cells, or bodily fluids to diagnose pathologic disorders or help with determining the prognosis of a disorder's progression. Examining tissue samples of solid tumors, microscopes have been the most commonly used tool in this field before the introduction of digital scanners \cite{microscopy_in_pathology} to convert glass slides with mounted processed and fixed tissue samples into a digital image. This digitization introduced a new concept to the field commonly referred to as \emph{digital pathology} \cite{digital_pathology}. While digital pathology is the general term used to describe integrating innovative digital tools into pathology, computational pathology describes a field that uses computational techniques and algorithms to analyze and interpret pathology data with the goal of improving the diagnosis, prognosis, and treatment of diseases \cite{computational_pathology}. 

As tissue glass slides constitute a major source of information in pathology, whole slide imaging and whole slide image (WSI) analysis comprise a major and new subset of computational pathology. The current market of digital scanners is very diverse with most of the commercially available digital scanners being able to capture whole slide images at magnifications as high as 40X, generating extremely large image files \cite{wsi_imaging_hardware}. While storing and leveraging the information in such a huge amount of data can be challenging, the stored retrospective data can be effectively used to introduce assistive tools into digital pathology increasing the speed and accuracy of pathologists hence reducing the workload. Artificial intelligence (AI) has shown promise in utilizing mass medical archives.

Deep learning uses artificial neural networks to learn complex patterns in any type of data \cite{deep_learning_review}. In contrast to conventional machine learning which mainly uses structured data, e.g., tabular data, the so-called deep models (deep artificial neural networks) can take in unstructured data including text, sound, and image, without pre-processing or pre-extraction of features. This makes deep networks extremely useful in real-life applications, including medicine. Deep models are mainly composed of multiple layers of connected artificial neurons. The general workflow of training and using a deep model in a supervised manner includes presenting the model with enough samples so the model can be properly trained to perform the desired task on new cases \cite{deep_learning_review}.

It has been shown that searching in an indexed dataset with image-level labels may provide a base for computational consensus \cite{atlas, yottixel, like_yottixel}. An index dataset can be called an ``atlas'' of medical information where previously established knowledge is stored and can be retrieved for patient matching. In such an atlas, newly encountered cases, i.e., new patients, are indexed using the same methods and are matched against all the atlas cases to find the most similar patients. Diagnosis, staging, prognosis, and any other decision or prediction can be then inferred from the top similar atlas patients for which all outcomes are known. 

Indexing visual content as generating a vector is the major task for atlas creation. While deep networks are mainly used as classifiers to provide definite labels on the input data, it has been shown that the numerical representations of some pre-decision layers of the network, also called embeddings or deep features, contain important information about the content of the input data \cite{deep_features_lowe, deep_features_babenko}. The concept of image retrieval in an atlas using deep features has been used in both radiology and pathology \cite{yottixel,like_yottixel, embd_cbir_neuroradiology}. Any deep network can be used to obtain deep features, but the architecture of the network and the data it has been trained on can greatly affect the representativeness of the features for the task at hand. For instance, KimiaNet is a dense convolutional neural network fine-tuned using all diagnostic histopathology images from the TCGA repository \cite{kimianet}. Vision transformers have also shown a promising ability to capture the details in a given image \cite{vision_transformer, dino1}.

While deep networks are widely used for generating deep features from images, they also impose limitations regarding image size. State-of-the-art graphical processing units (GPUs) which are commonly used in deep learning can process up to a certain size of the image in each iteration. Therefore, any whole slide image (WSI) processing requires the image to be broken down into many smaller images, also known as \emph{patches} \cite{dl_in_wsi_review}. While patching gigapixel WSIs is a crucial step in computational pathology, one WSI can generate several hundred and even several thousands of patches of ordinary size, say 224 by 224 or 512 by 512 pixels, depending on the size of the WSI, magnification, and the biopsy type. One approach would be to feed all the patches into the deep network and use the output for the downstream tasks. Such \emph{brute force} approach would require massive computational power, a factor that would limit the adoption of the technology.

Processing WSIs is a difficult problem that requires a \emph{divide \& conquer} approach. Patching is generally the divide and extracting deep features is the conquer part. Various methods have been proposed to select a smaller subset of WSI patches as representative ones for downstream tasks. One approach as described in Yottixel is clustering patches based on color distribution or RGB histogram and proximity of patches in WSI to select a diverse but smaller set of patches from all morphological structures within a WSI \cite{yottixel}. This is called a \emph{mosaic} of WSI. The Yottixel mosaic concept is simple and reliable and has been adopted by other works \cite{wang2023retccl,sikaroudi2023comments}. Still, patch selection remains one of the open fields for research because it can greatly affect the performance of the downstream analyses. Yottixel's mosaic ignores normal histology and selects normal and malignant patches based on their frequency in the clustering space.

As pathologists have been trained to recognize normal histologies and disregard them when focusing on the pathology, it can be helpful for a better patching strategy to remove normal histology specific to the site. One way to formulate this task is by integrating anomaly detection into the patching scheme. Provided WSIs depicting entirely normal tissue are available, one-class classifiers \cite{one_class_survey} could learn the typical features of normal histology. The trained classifiers can then accompany the mosaic generation, i.e., the patching, to remove normal patches.  

In this paper, we propose and validate an ``atlas of normal tissue'' solely using samples of WSIs obtained from normal tissue biopsies. The proposed method leverages a weakly supervised multiple-instance learning method using one-class classifiers to exclude the normal patches and focus on clinically important parts of any given WSI. We argue that this method will reduce the number of patches selected, hence removing redundancy, from each WSI while maintaining the representativeness of the patches for downstream tasks.
\section*{Experiments Setup}
\textbf{Datasets.} Two datasets from two different organs were used to test the proposed atlas of normal atlas and the one-class classifier for normal patch detection. 
\begin{itemize}
    \item One dataset included a total number of 660 skin tissue WSIs of patients diagnosed with \textbf{cutaneous squamous cell carcinoma} (cSCC) which were pulled from the internal Mayo Clinic database (REDCap). The cases were first identified through enterprise-wide internal search engines which identified patients with a histopathologic diagnosis of cSCC or metastatic cSCC from pathology reports of archived tissue specimens. These cases subsequently underwent chart review to confirm the primary tumor as part of the inclusionary criteria. Samples were either taken at Mayo Clinic locations (Minnesota, Arizona, Florida) or at outside facilities and shared with Mayo Clinic for consultation. The tissue may be from either biopsy (punch, shave, etc.) or from subsequent excision. All tissue included in the pilot underwent both initial review and re-review by a dermatopathologist at one of the Mayo Clinic sites. The purpose of the re-review was to confirm cSCC, tumor characteristics (depth, differentiation, PNI, etc.), and tumor stage, and to select cases most appropriate for scanning and sequencing.

    Selected cases included 386 well-differentiated, 100 moderately differentiated, and 67 poorly differentiated cSCC. There were also 107 normal WSIs selected to represent the normal skin tissue. The only associated label used is the degree of differentiation or being normal. A total number of 10 slides were randomly selected from well-differentiated and poorly differentiated cases for internal validation of the normal atlas at the patch level. A pathologist selected the regions with the most prominent abnormal morphology in these slides. The annotations were later used to generate normal and abnormal patches for validation purposes only.

    \item The second collected dataset included 21 breast tissue WSIs obtained from 8 patients with no prior diagnosis of breast cancer at the Mayo Clinic. For validation purposes, 78 WSIs from cases with lobular carcinoma of the breast and 354 WSIs from cases with ductal carcinoma of the breast were obtained from the TCGA Research Network. These were all the diagnostic WSIs available on the repository for the two selected medical conditions at the time of collecting the data. Tissue samples from cases with no prior malignancy and no treatment at the time of sampling were selected to imitate the diagnostic pipeline of a deployed model.
\end{itemize}

\textbf{Normal Atlas Creation.} The idea of the normal atlas relies upon leveraging anomaly detection algorithms used in machine learning to \emph{detect outliers} in a dataset. One-class classifiers are best suited for this task. Due to the complex nature of the data, we selected Isolation Forest \cite{liu2008isolation} and One-Class SVM \cite{one_class_svm} as the abnormality detection mechanisms in this study. Both methods were presented with a series of deep features obtained from normal patches. The fitted classifiers were then used to classify normal and abnormal in a given set of deep features they had not seen before. Data preprocessing included tissue segmentation, patching, and color normalization. The tissue segmentation was carried out using an in-house trained U-NET segmentation model. Patching was done at 20X magnification with a patch size of 1024 by 1024 and an overlap of 30 percent between adjacent patches, both in height and width. A minimum of 75 percent patch area coverage was selected as the criteria for excluding patching with insufficient tissue. All patches were separately fed to three different deep networks to obtain the deep features. These networks included one convolutional neural network with the architecture of DenseNet121 specifically trained on pathology patches named KimiaNet \cite{kimianet}, one pre-trained vision transformer using a method named DINO, and another convolutional neural network with the architecture of ResNet50 trained using the same DINO approach \cite{dino1}.

\begin{figure*}[ht!]
    \centering
    \includegraphics[width=\linewidth]{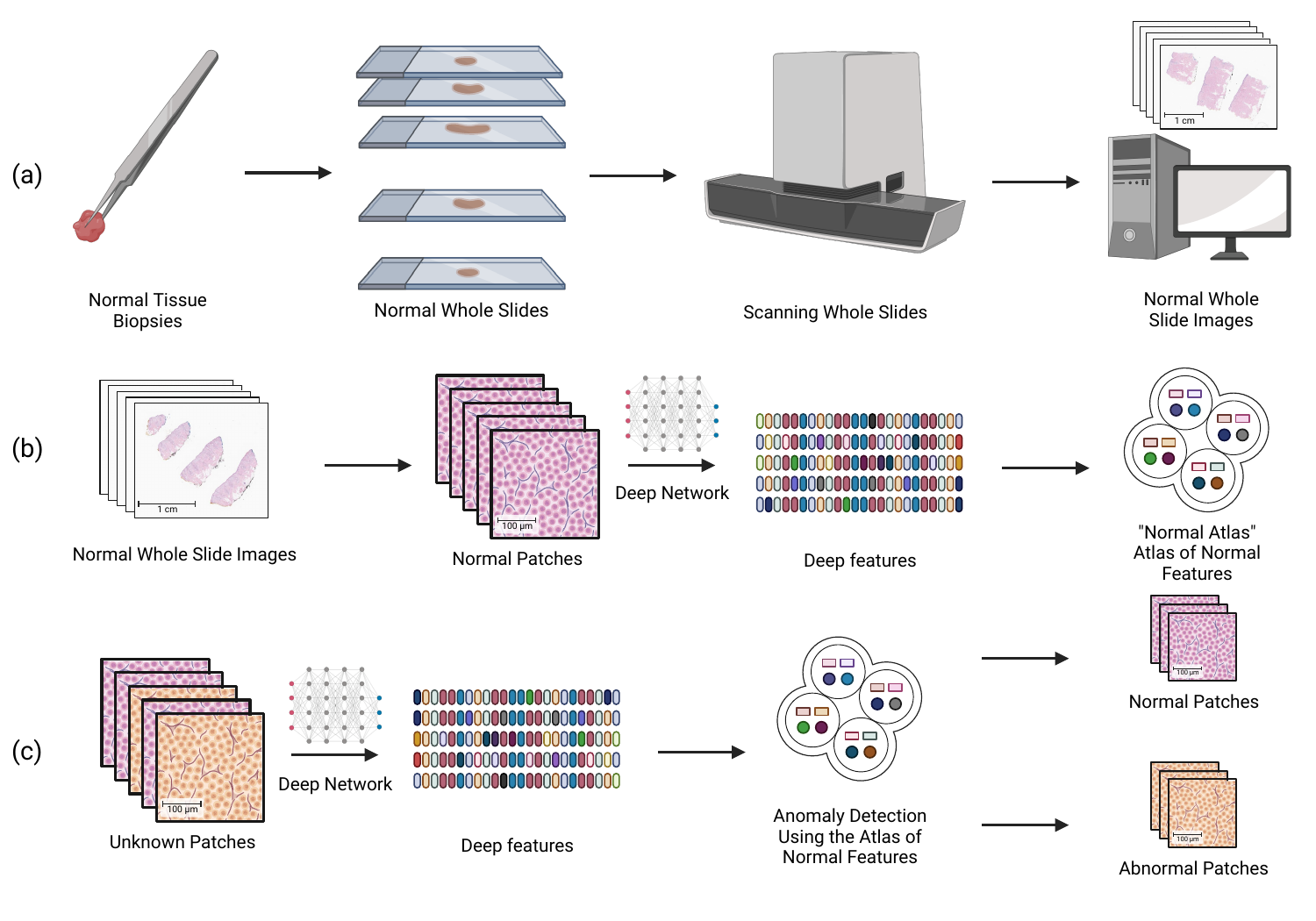}
    \caption{The general workflow of creating the normal atlas: (a) Normal WSIs are generated by scanning slides of known normal tissue biopsies, (b) Deep features of normal WSIs are obtained by passing the patches from each WSI through a deep network; the resulting deep features are stored in the normal atlas, (c) The constructed normal atlas is used to differentiate normal vs abnormal deep features from unknown patches obtained using the same preprocessing and feature extraction method.}
    \label{fig:1_normal_atlas_creation}
\end{figure*}

When training the one-class classifiers with the normal data, the sample was assumed to be contaminated with some abnormal data. In the context of normal WSIs, this may include artifacts, tissue folds, stain sedimentation, pen markings, and out-of-focus slide areas. Both Isolation Forest and one-class SVM use methods to incorporate this assumption. One-class SVM utilizes a factor called $\nu$ to set an upper bound on the fraction of margin errors and a lower bound on the fraction of support vectors relative to the total number of training examples \cite{nu_factor_svm}. We set $\nu=5\%$ of the total cases according to the expert's opinion and let the Isolation Forest classifier deduct this parameter automatically according to the method described in the original paper \cite{liu2008isolation}. The pseudocode of creating and using the normal atlas is outlined in algorithms \ref{algorithm:creating_normal_atlas} and \ref{algorithm:using_normal_atlas}. The pipeline of creating the normal atlas is also visually shown in Figure \ref{fig:1_normal_atlas_creation}. We used the Python library SciKit Learn for the implementation of Isolation Forest and One-Class SVM \cite{scikit-learn}.

\begin{algorithm}[ht!]
    \caption{Pseudocode for creating the normal atlas given a set of normal WSIs $I_{N}$}
    \label{algorithm:creating_normal_atlas}
    \begin{algorithmic}
        \State Set $I_{N}$ (set of normal WSIs)
        \State Set $W$ (deep network)
        \State Set $M$ (unfitted instance of one-class classifier)
        \State Set $s$ (patch size)
        \Procedure{CreateNormalAtlas}{$I_{N}$}
            \State $F \gets \emptyset$
            \Comment{Instantiating $F$ as an empty set of normal patch deep features}
            \For{$I \in I_{N}$}
                \State $T \gets$ TissueSegmentation($l$)
                \Comment{Extract the tissue regions $T$}
                \State $P_{I} \gets$ Patching($I, s$)
                \Comment{Perform patching for patch size $s$ and storing results in $P_{I}$}
                \State $P_{T} \gets T \cap P_{I}$
                \Comment{Isolate patches containing tissue regions as $P_{T}$}
                \State $F_{I} \gets$ DeepNet($P_{T}$, $W$)
                \Comment{Extract the features using deep network $W$}
                \State $F \gets F_{I} \cap F$
                \Comment{Append deep features from WSI $I$ to normal patch deep features $F$}
            \EndFor
            \State $M_{F} \gets$ FitOneClassClassifier($F$, $M$)
            \Comment{Fitting the one-class classifier using the patch features}
            \State \textbf{return} $M_{F}$
        \EndProcedure
    \end{algorithmic}
\end{algorithm}

\begin{algorithm}[ht!]
    \caption{Pseudo-code for using the normal atlas given a patch $P$ and fitted one-class classifier $M_{F}$}
    \label{algorithm:using_normal_atlas}
    \begin{algorithmic}
        \State Set $M_{F}$ (fitted one-class classifier)
        \State Set $P$ (query patch)
        \State Set $W$ (deep network)
        \Procedure{UseNormalAtlas}{$P, M_{F}$}
            \State $F \gets$ DeepNet($P$, $W$)
            \Comment{Extract the features of patch $P$ using deep network $W$}
            \State $N \gets$ DetectAnomaly($F, M$)
            \Comment{Deciding whether the deep features $F$ are normal or abnormal using the fitted classifier $M_{F}$}
            \If{$N$ is normal}
                \State \textbf{return} $True$
                \Comment{Returning True means that the given patch $P$ is normal}
            \Else
                \State \textbf{return} $False$
                \Comment{Returning False means that the given patch $P$ is not normal}
            \EndIf            
        \EndProcedure
    \end{algorithmic}
\end{algorithm}

\begin{figure*}[ht!]
    \centering
    \includegraphics[width=\linewidth]{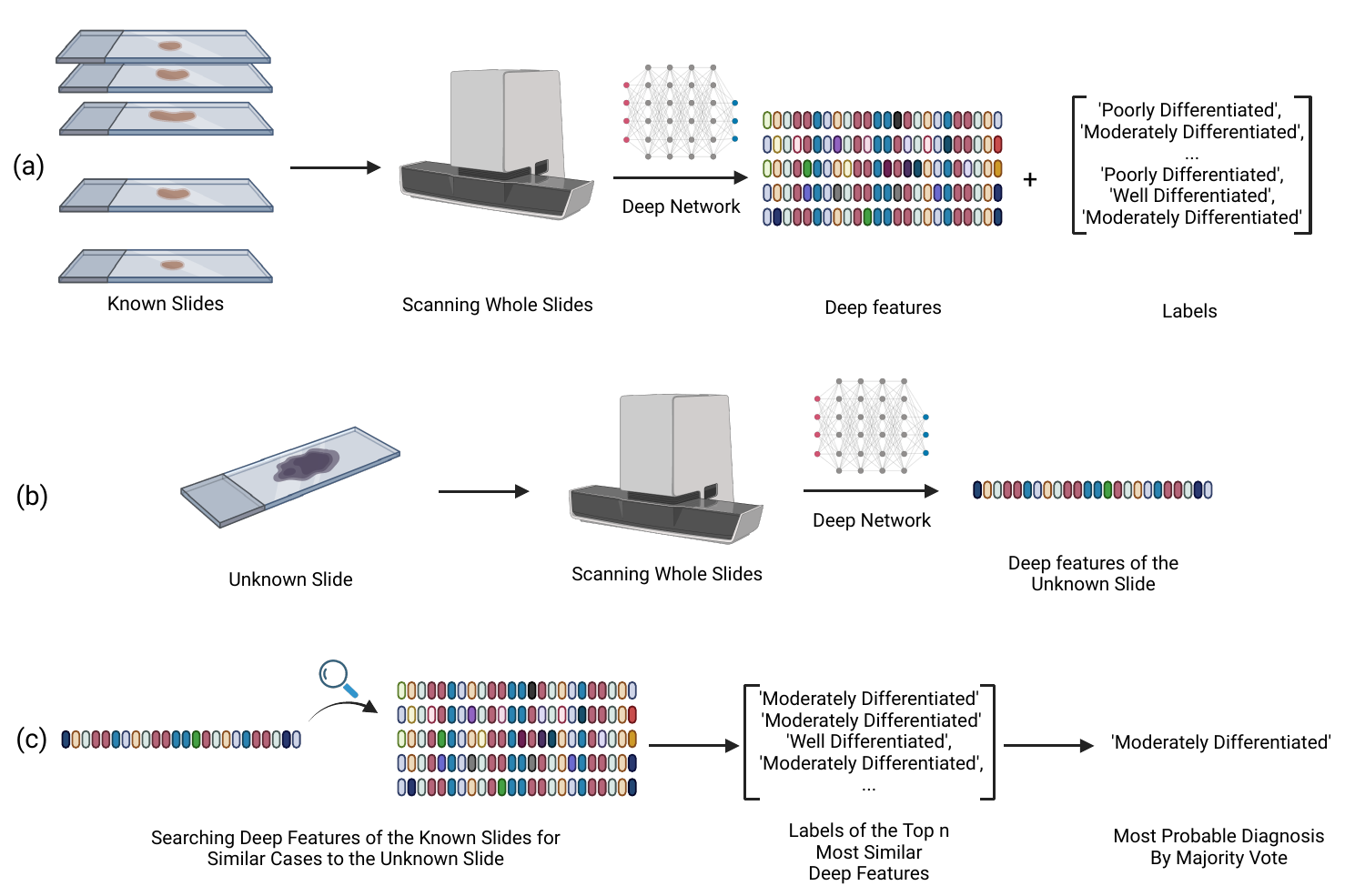}
    \caption{General workflow of index and search: (a) Deep features of known slides are obtained via passing the preprocessed patches through a deep network and are stored with their corresponding labels, (b) The deep features of a new (diagnostically unknown) slide are obtained through the same process, and (c) Stored deep features of the known slides are searched to find the closest matches to the newly generated deep features to extract the most probable label through a majority vote among the top-n closest labels.}
    \label{fig:2_index_and_search_workflow}
\end{figure*}

\textbf{Index and Search Workflow.} To demonstrate the role of removing the normal patches from the WSI, we have used the general workflow for indexing and searching of WSIs for retrieval of similar cases as previously explained in Yottixel with some minor adjustments \cite{yottixel}. The tissue segmentation and patching follow the same method used to obtain the normal patches. To reduce the computational and storage cost, a subset of patches is then selected from each WSI using the patch selection method of Yottixel \cite{yottixel}. This method involves clustering patches into 9 clusters using a histogram of their RGB values and then selecting 15 percent of the patches from each cluster in a spatially homogenous manner. This results in selecting almost 15 percent of the total patches from each WSI, collectively called mosaic. The deep features for the patches of each WSI are used as an indexed version of that WSI. The median of minimum one-to-one Euclidean distances between all the patches of the two WSI is used to find the most similar indexed WSIs to any given WSI. The same approach for obtaining similarity is used in Yottixel with the difference being the Hamming distance for barcoded features \cite{yottixel} instead of the Euclidean between deep features which was used in our study. The schematic workflow of indexing and searching WSIs is shown in Figure \ref{fig:2_index_and_search_workflow}.

\textbf{Internal and External Validation.} The performance of the normal atlas at the patch level was tested by obtaining normal and abnormal patches from 10 randomly selected cSCC test slides which were annotated by a pathologist. Having a minimum of 50 percent surface area of the patch annotated as the abnormal region was set as the threshold to label the patch as \emph{abnormal}. The rest of the patches were considered normal. The patches were obtained in the same size and magnification as the original datasets used, i.e., 1024 by 1024 at 20X. This process yielded 6479 normal and 9352 abnormal patches. Precision, recall, and F1 scores were selected as performance metrics for this classification task.

Three different setups of indexing and search pipeline were examined to test the effect of the normal atlas at the WSI level. The base setup included the original indexing and search pipeline for the WSIs as described in the Yottixel paper \cite{yottixel}. Two other setups were almost the same with one added step of excluding normal patches using an atlas of normal tissue after applying the Yottixel's mosaic to acquire patches. One setup used the Isolation Forest as the method of anomaly detection and another one used the one-class SVM. The resulting patches were also used to perform indexing and search in the same manner as the base setup. WSIs with all patches excluded as normal patches were labeled as normal. Performance of the indexing and search pipelines were measured using a leave-one-patient-out validation for all included patches by taking the output attached to top-1 and a majority vote among top-3 and top-5 most similar WSIs.
Performance metrics such as precision, recall, and F1 score were calculated for each label inference. The overall performance of each configuration was also reported using the weighted F1 score of all the labels in the dataset according to their number of occurrences.
\section*{Results}
\begin{figure*}[ht!]
\centering
\includegraphics[width=\textwidth,height=\textheight,keepaspectratio]{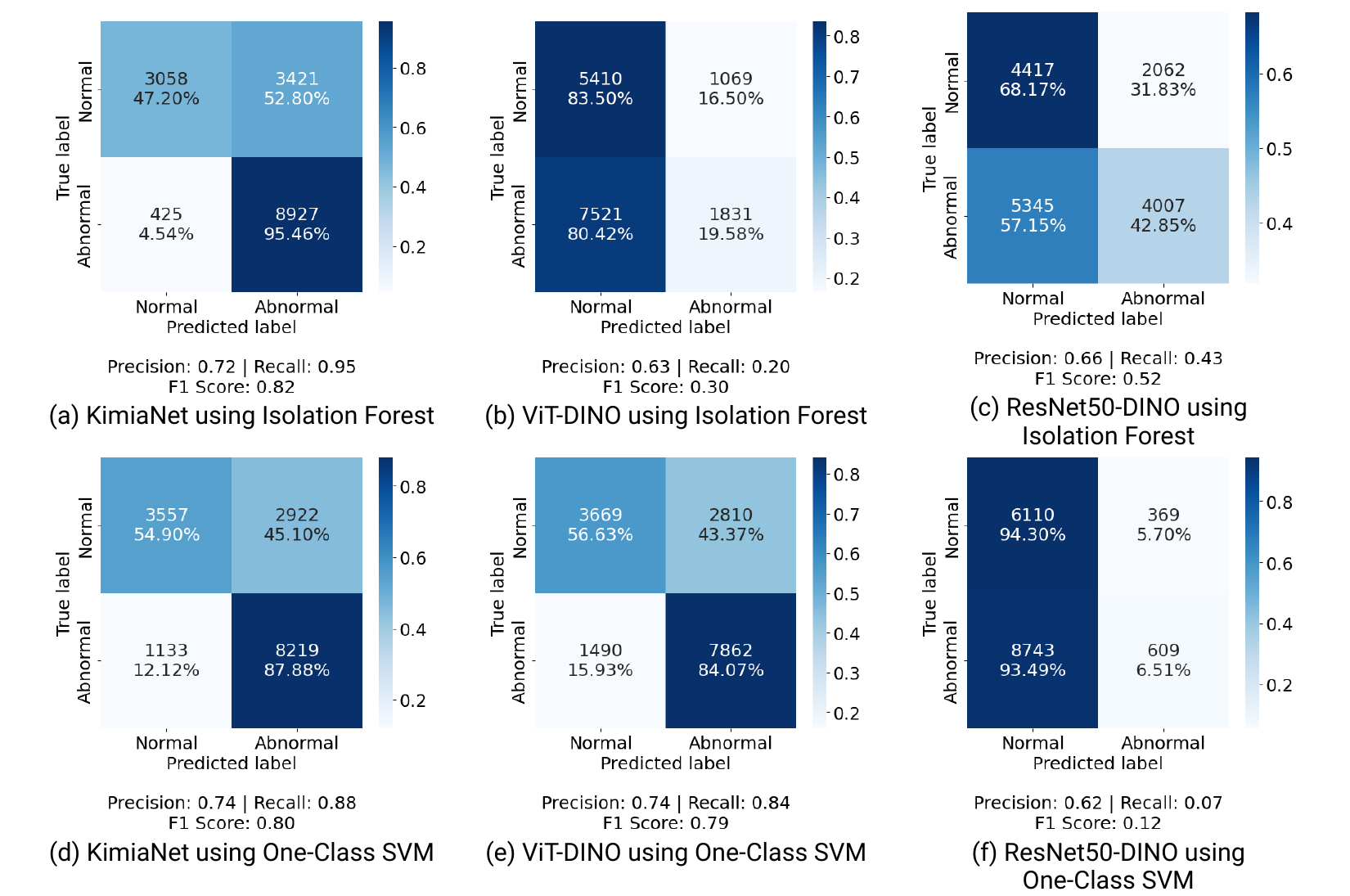}
\caption{Normal atlas performance at patch-level demonstrated using validation results for 10 manually annotated WSIs by a pathologist shown as confusion matrices using normal atlases established using Isolation Forest (a,b,c) and one-class SVM (d,e,f) on deep features extracted by KimiaNet (a,d), ViT-DINO (b,e), and ResNet50-DINO (c,f).}
\label{fig:4_normal_validation_skin}
\end{figure*}

\begin{figure}[ht!]
\centering
\includegraphics[width=\columnwidth,height=0.7\textheight,keepaspectratio]{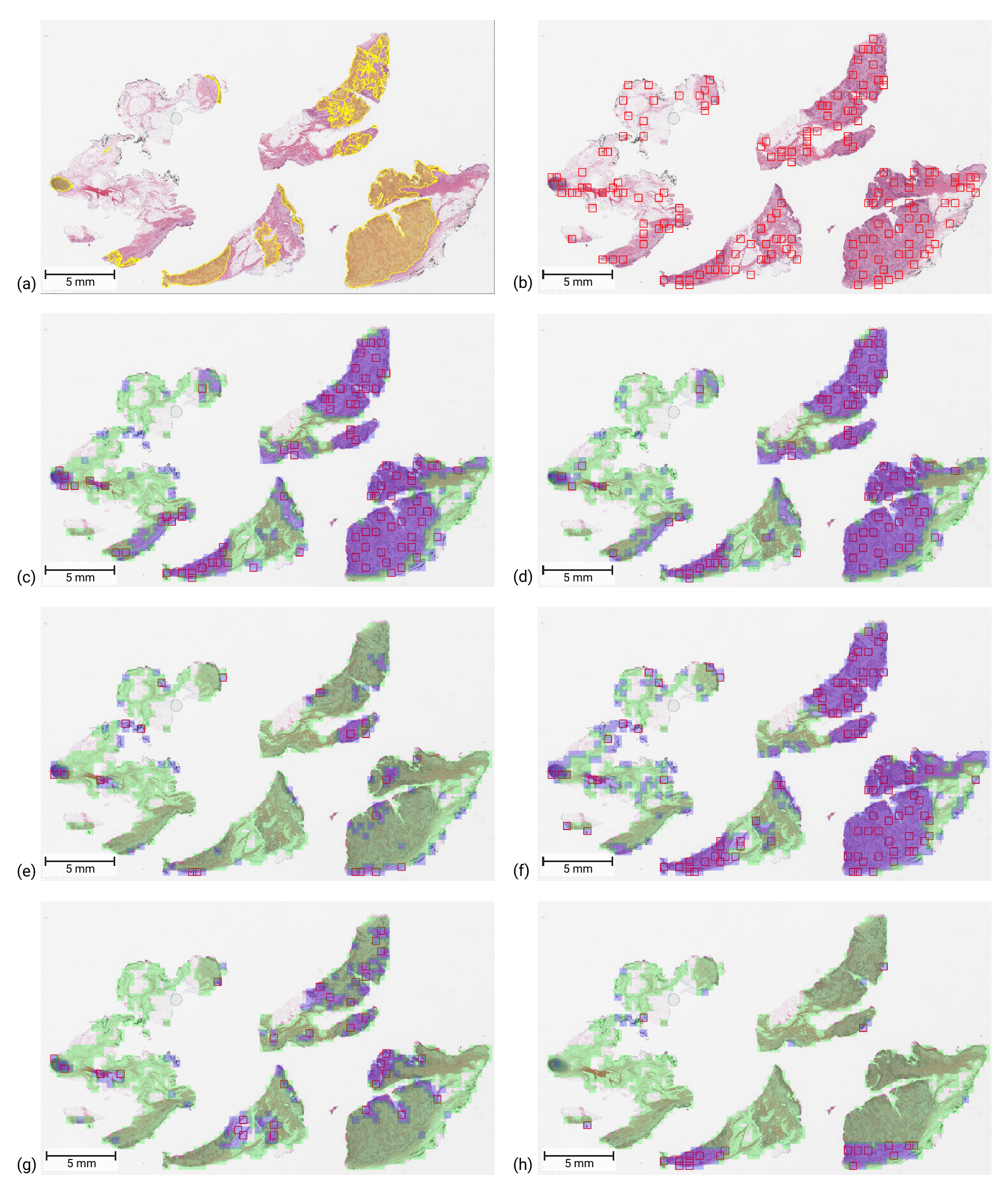}
\caption{Visual results in a cSCC case: (a) Pathologist ground-truth, (b) Yottixel's mosaic, (c) KimiaNet and Isolation Forest, (d) KimiaNet and one-class SVM, (e) ViT-DINO and Isolation Forest, (f) ViT-DINO and one-class SVM, (g) ResNet50-DINO and Isolation Forest, (h) ResNet50-DINO and one-class SVM. Red boxes show the patches selected for indexing and search within each WSI.}
\label{fig:5_visual_example_skin}
\end{figure}

\begin{figure}[ht!]
\centering
\includegraphics[width=\columnwidth,height=0.7\textheight,keepaspectratio]{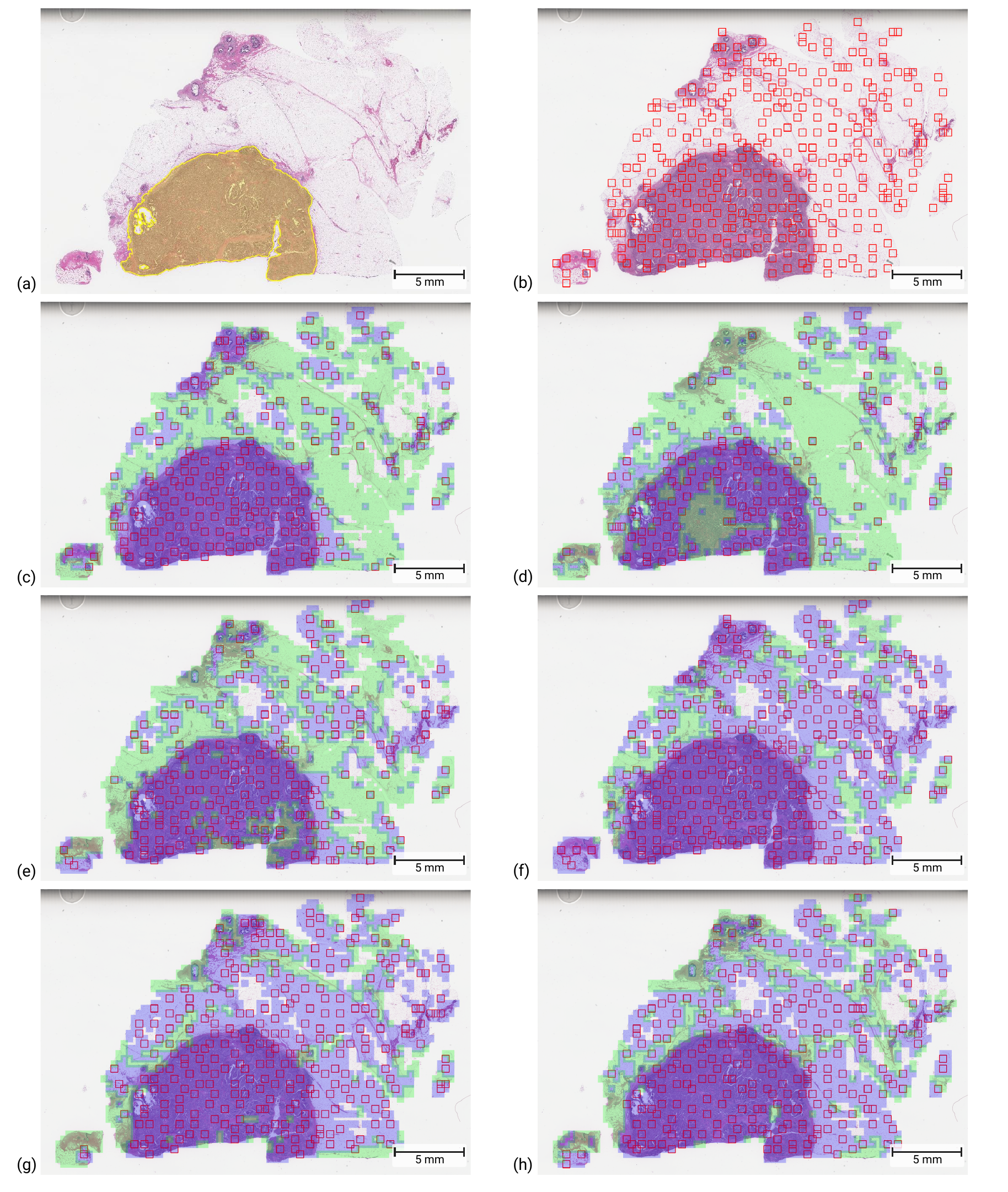}
\caption{Visual results of a breast carcinoma case: (a) Pathologist ground truth, (b) Yottixel's mosaic, (c) KimiaNet and Isolation Forest, (d) KimiaNet and one-class SVM, (e) ViT-DINO and Isolation Forest, (f) ViT-DINO and one-class SVM, (g) ResNet50-DINO and Isolation Forest, and (h) ResNet50-DINO and one-class SVM. Red boxes show the patches selected for indexing and search within each WSI.}
\label{fig:7_visual_example_breast}
\end{figure}

\textbf{Normal Atlas Patch-Level Validation.} The patch-level internal validation for the normal atlases was carried out using the 10 WSIs which were annotated for abnormal and normal regions by a pathologist. The performance varied based on the deep network used to obtain the deep features and the anomaly detection method used. KimiaNet's features showed the best performance both using Isolation Forest and one-class SVM, with F1 scores of 0.82 and 0.80, respectively. Among the networks trained with natural images, the DINO-trained vision transformer combined with one-class SVM showed the best results with an F1 score of 0.79. The detailed patch-level validation results for normal skin atlases are shown in Figure \ref{fig:4_normal_validation_skin}. One visual example of how patch-level classification of different deep networks and anomaly detection algorithms comply with pathologist annotations is shown in Figure \ref{fig:5_visual_example_skin} for a cSCC WSI and in Figure \ref{fig:7_visual_example_breast} for a breast cancer WSI.

\begin{figure}[ht!]
\centering
\includegraphics[width=\columnwidth,height=\textheight,keepaspectratio]{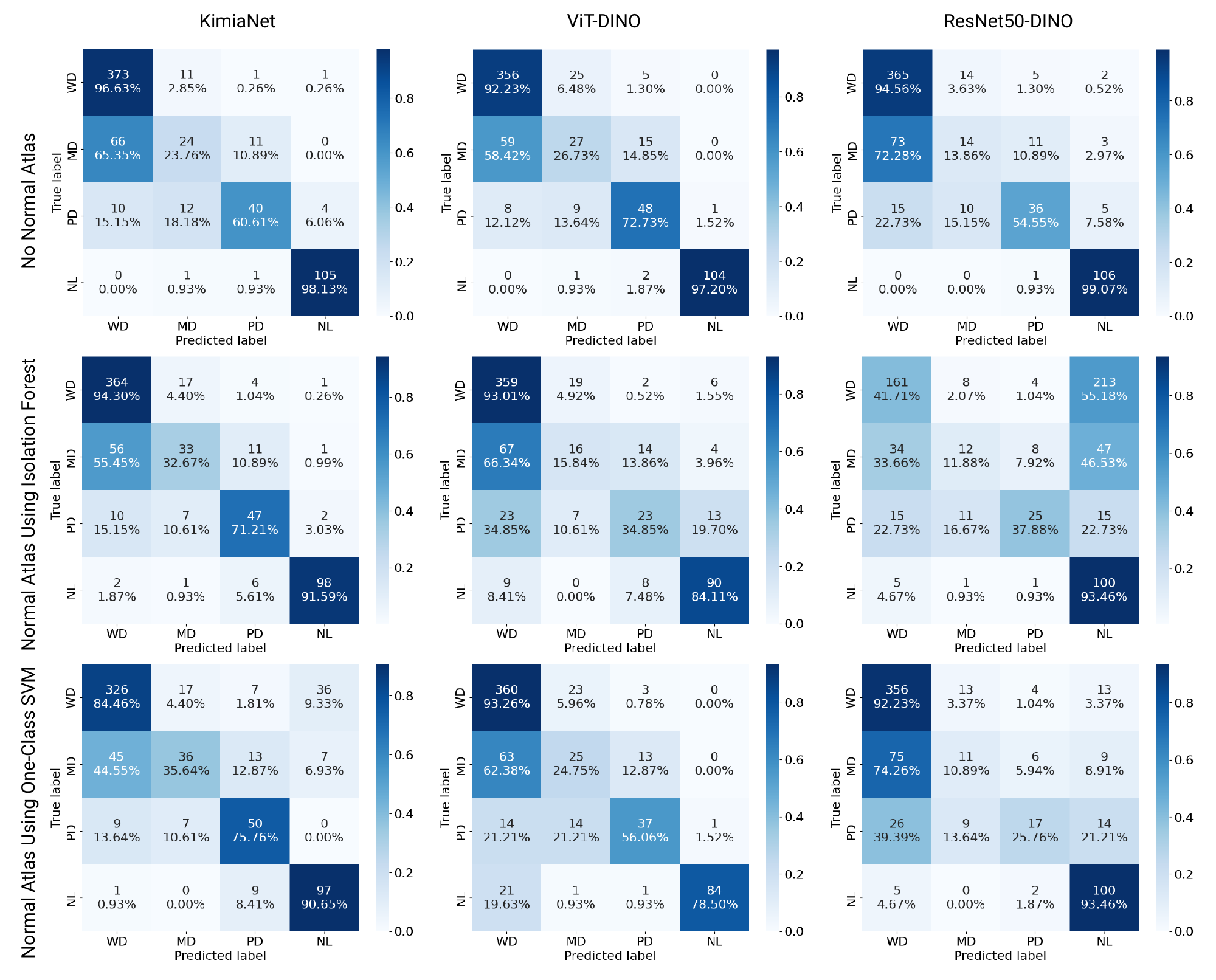}
\caption{Indexg and search confusion matrices for leave-one-patient-out validation for top-5 consensus predictions in skin dataset. Columns represent the indexing and search using deep features from KimiaNet, DINO-trained vision transformer, and DINO-trained ResNet50, respectively. Rows represent results in three experimental setups using no normal atlas, using a normal atlas with Isolation Forest, and using a normal atlas with one-class SVM, respectively.}
\label{fig:3_index_and_search_results_skin}
\end{figure}

\begin{figure}[ht!]
\centering
\includegraphics[width=\columnwidth,height=\textheight,keepaspectratio]{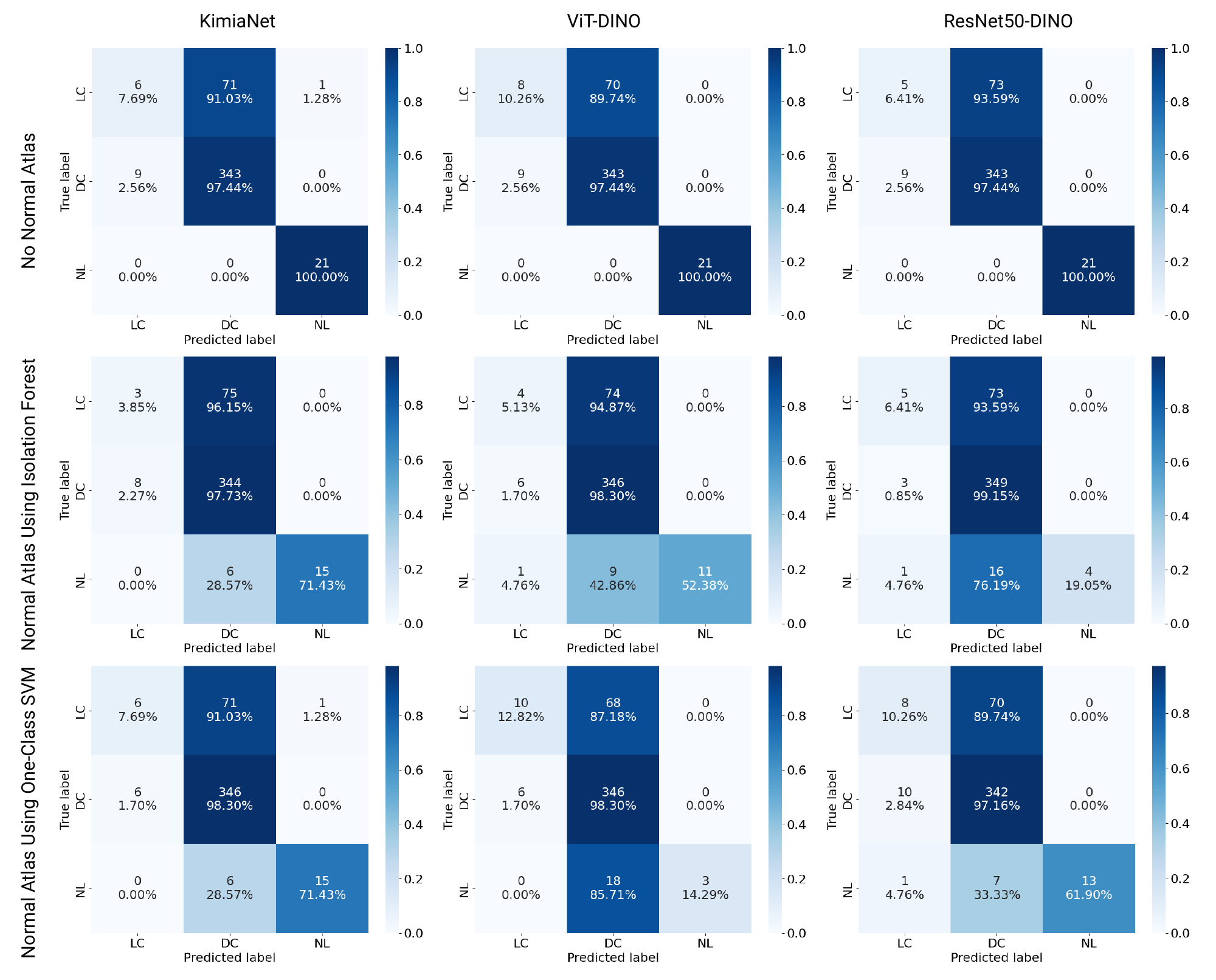}
\caption{Indexing and search confusion matrices for leave-one-patient-out validation for top-5 consensus predictions in breast dataset. Columns represent the indexing and search using deep features from KimiaNet, DINO-trained vision transformer, and DINO-trained ResNet50, respectively. Rows represent results in three experimental setups using no normal atlas, using a normal atlas with Isolation Forest, and using a normal atlas with one-class SVM, respectively.}
\label{fig:6_index_and_search_results_breast}
\end{figure}

\begin{figure}[ht!]
\centering
\includegraphics[width=\columnwidth,height=0.7\textheight,keepaspectratio]{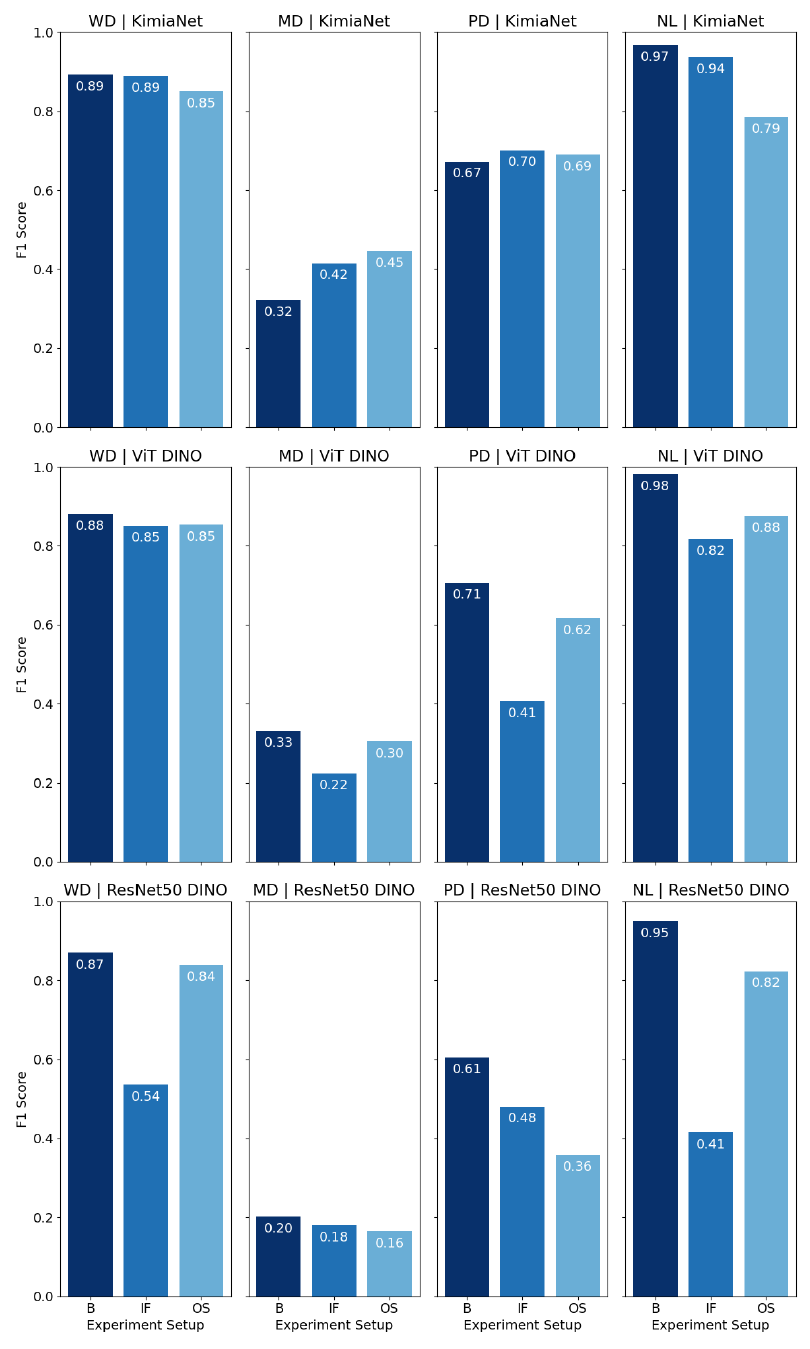}
\caption{Indexing and search F1 score for leave-one-patient-out validation for top-5 consensus predictions in skin dataset plotted for each class, deep network, and experimental setup separately (WD = well-differentiated, MD = moderately differentiated, PD = poorly differentiated, NL = normal, B = base configuration of using no atlas, IF = using normal atlas based on Isolation Forest, OS = using normal atlas based on one-class SVM).}
\label{fig:8_top_5_f1_score_skin}
\end{figure}

\begin{figure}[ht!]
\centering
\includegraphics[width=0.7\columnwidth,height=0.737\textheight,keepaspectratio]{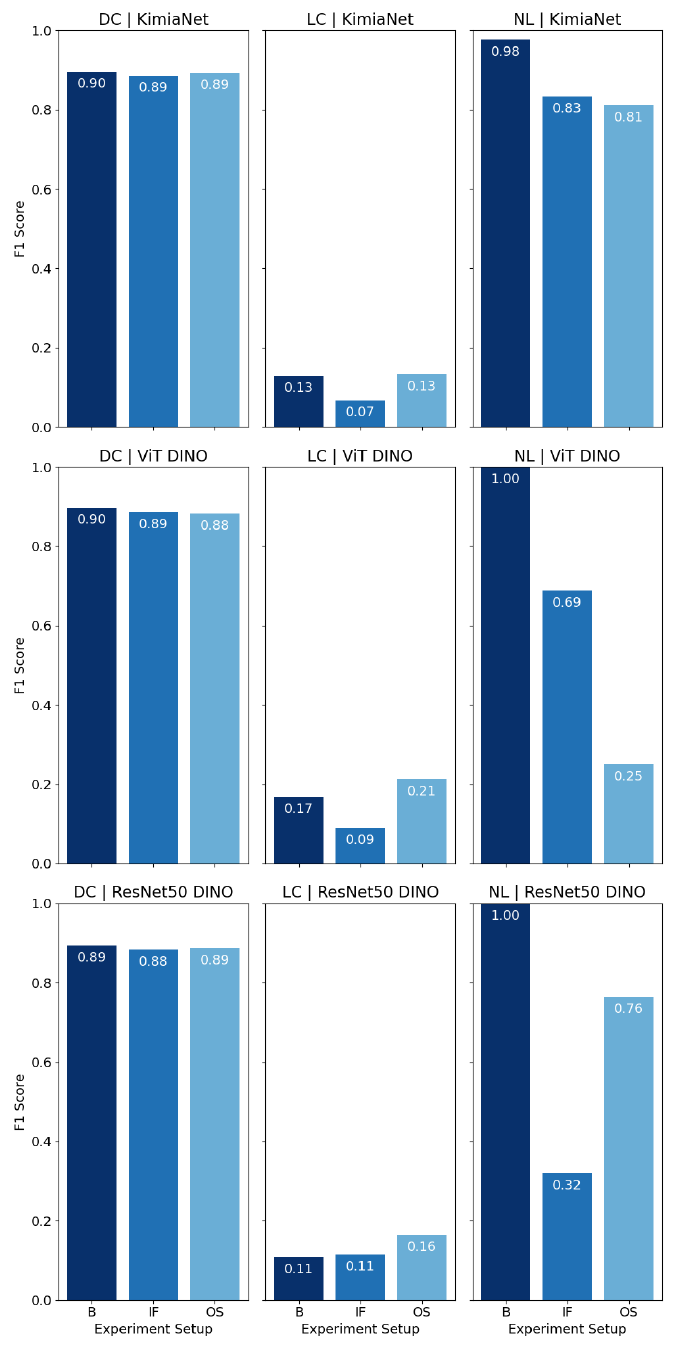}
\caption{Indexing and search F1 score leave-one-patient-out validation for top-5 consensus predictions in breast dataset plotted for each class, deep network, and experimental setup separately (DC = infiltrating ductal carcinoma, LC = lobular carcinoma, NL = normal, B = base configuration of using no atlas, IF = using normal atlas based on Isolation Forest, OS = using normal atlas based on one-class SVM).}
\label{fig:9_top_5_f1_score_breast}
\end{figure}

\textbf{Indexing and Search Results.} Indexing and search were carried out on both datasets of cSCC and breast cancer WSIs using the three previously described experiment setups. The detailed performance results of the indexing and search experiments are reported in Tables \ref{tab:metrics_table_skin} and \ref{tab:metrics_table_breast} in the supplementary material section. As the confusion matrices for the leave-one-patient-out classification depicted in Figure \ref{fig:3_index_and_search_results_skin} for the skin dataset show the two mostly mislabeled classes are 'well-differentiated' and 'moderately differentiated'. While this is true for the indexing and search using the deep features from all three deep networks, KimiaNet (trained specifically on pathology images \cite{kimianet}) shows the best performance. The classification, regardless of the network used to obtain the deep features, shows almost the same level of performance in all three setup configurations of using no normal atlas, using a normal atlas with Isolation Forest, and using a normal atlas with one-class SVM. To show this in the skin dataset, the F1 score for different experimental setups of the top-5 consensus is illustrated in Figure \ref{fig:8_top_5_f1_score_skin}. Confusion matrices of the indexing and search leave-one-patient-out validation for the breast dataset are shown in Figure \ref{fig:6_index_and_search_results_breast}. Indexing and searching for breast carcinoma cases using deep features from all three models also show suboptimal performance for lobular carcinoma and infiltrating ductal carcinoma. However, the performance stays almost the same in three setups of not using the normal atlas, using one with Isolation Forest, and using one with a one-class SVM. This is shown in the  F1 scores plot of the top-5 consensus of the corresponding experiments in Figure \ref{fig:9_top_5_f1_score_breast}. While the performance of the classification using the indexing and search stays almost the same in both skin and breast datasets, the number of total patches selected in the whole dataset to represent the WSIs is reduced by at least 12\% and at most 86\% of the original number of patches as shown in Table \ref{tab:patch_numbers}. The exact level of patch reduction depends on the deep network used to obtain the deep features and on the anomaly detection algorithm used in the normal atlas.

\begin{table*}[ht!]
\centering
\caption{Number of patches used for indexing and search in three different experimental setups including no normal atlas, normal atlas with Isolation Forest, and normal atlas with one-class SVM for skin and breast WSIs. For each setup, the numbers have been reported for deep features obtained from all three networks. The percentage inside the parenthesis for each cell shows the number of patches excluded compared to the base configuration, i.e., no normal atlas used, in each experiment.}
\label{tab:patch_numbers}
\begin{tblr}{
  width = \linewidth,
  colspec = {Q[69]Q[198]Q[223]Q[223]Q[223]},
  cells = {c},
  cell{1}{1} = {c=2}{0.267\linewidth},
  cell{2}{1} = {r=3}{},
  cell{5}{1} = {r=3}{},
  hline{1-2,5,8} = {-}{},
}
\textbf{Experiment Setup} &  & {\textbf{KimiaNet}\\\textbf{Patches (\% Reduction)}} & {\textbf{ViT DINO}\\\textbf{Patches (\% Reduction)}} & {\textbf{ResNet50 DINO}\\\textbf{Patches (\% Reduction)}}\\
Skin & No Normal Atlas & 43830 (0\%) & 43858 (0\%) & 43829 (0\%)\\
 & {Normal Atlas\\Using Isolation Forest} & 21799 (50\%) & 7698 (82\%) & 8820 (80\%)\\
 & {Normal Atlas\\Using One-Class SVM} & 14343 (67\%) & 22254 (49\%) & 6237 (86\%)\\
Breast & No Normal Atlas & 112090 (0\%) & 112090 (0\%) & 112090 (0\%)\\
 & {Normal Atlas\\Using Isolation Forest} & 97657 (13\%) & 59879 (47\%) & 65343 (42\%)\\
 & {Normal Atlas\\Using One-Class SVM} & 89810 (20\%) & 99028 (12\%) & 65983 (41\%)
\end{tblr}
\end{table*}

\section*{Discussion}
While most of the computational pathology tasks are designed to classify or detect the presence of pathological lesions in gigapixel WSIs, the presence of computationally redundant normal tissue in WSIs is often ignored. While a large body of scientific work has been done in the field of WSI retrieval systems to address the numerical representation of the WSI \cite{fisher_vector, kimianet, cbir_in_pacs, contarstive_representation_learning} and WSI search engines \cite{yottixel, like_yottixel, wsi_retrieval, sikaroudi2023}, no study has addressed the effect of normal tissue in the indexing and search pipeline. Therefore, removing normal tissue is not an established practice in this field. While some works have used normal tissue exclusion as part of their workflow \cite{used_normal_exclusion_poster}, the potential effect of redundant normal tissue in WSI retrieval is an unknown area. In this work, we tested the use of two anomaly detection algorithms as a means of excluding abnormal patches from Yottixel's indexing and search \cite{yottixel}. Our methods included no WSI-level annotation and used multiple instance learning for known normal WSIs.

The urgent need for novel WSI patching is an important aspect of current works in the field of computational pathology. Patching involves dividing an image into smaller regions or patches and using these patches to represent the tissue in the image. There are several benefits to using tissue patching in computational pathology. First, it allows for the efficient analysis of large images, as the patches can be processed independently and in parallel. This can significantly reduce the computational resources required for image analysis. Second, WSI patching can improve the accuracy of algorithms by providing a more representative set of tissue patches. The main focus in most digital pathology tasks is to find the unique histopathologic fingerprint each WSI represents which is basically the diagnosis of the disease. This fingerprint might be unique to that WSI while normal (healthy) tissue constitutes a normal variation of cellular morphology and tissue structure that is common to all human histology in each primary site. Including these common features, namely normal histology, in indexing WSIs with relevant features, namely malignant lesions, introduces redundancy into the WSI retrieval workflow. 

Our results show that removing the areas with normal histology during the patching process helps with selecting a smaller set of representative patches for each WSI, therefore decreasing the computational and storage cost of the search engine, while maintaining the overall retrieval performance of the search engine.

As shown in Figure \ref{fig:8_top_5_f1_score_skin}, by using a one-class SVM to detect and exclude the abnormal patches in the cSCC dataset using features from KimiaNet, the classification performance for all three classes was either maintained at the same level or even improved. At the same time, as Table \ref{tab:patch_numbers} depicts, the number of patches for selected cases was reduced by 67\% after excluding the normal tissue. Contrary to this improvement, the performance for correctly classifying the normal cases decreases, with a decrease in the F1 score from 0.97 to 0.79. This drop can be attributed to the fact that excluding patches that look normal from a normal WSI will leave it with patches mainly containing artifacts and distortions which makes the indexing and search return mismatched results. In contrast to the improvement seen for the experiments with deep features obtained by KimiaNet, those done on deep features obtained using DINO-trained ViT and ResNet50 show rather mixed results.

The results for the breast dataset show almost the same pattern as the results for the cSCC dataset. As Figure  \ref{fig:9_top_5_f1_score_breast} shows, the classification performance for ductal carcinoma has shown minor changes regardless of the anomaly detection method selected and the network used to obtain the deep features. However, excluding normal tissue inflicts varying changes to the lobular carcinoma classification performance based on the network and anomaly detection method. Excluding normal tissue from breast WSIs also decreased the number of patches selected by 12\% to 47\% as depicted in Table \ref{tab:patch_numbers}. The same effect of decreased performance in normal cases for the cSCC dataset can also be observed in the breast dataset.

The changes in indexing and search performance seen in the skin dataset are also in concordance with the patch-level classification performance seen for annotated validation cases which is shown in Figure \ref{fig:4_normal_validation_skin}. The validation results show that using Isolation Forest and one-class SVM classifiers on deep features obtained from KimiaNet are the best combination with an F1 score of 0.82 and 0.80, respectively. The same configurations also showed the best results in indexing and search as shown in Figure \ref{fig:8_top_5_f1_score_skin}. Applying one-class SVM on DINO-trained ViT and Isolation Forest on DINO-trained ResNet50 also result in acceptable performance with an F1 score of 0.79 and 0.52, respectively. These two configurations also showed mediocre performance in indexing and search. Even though these two pairs of combinations showed good results, combining them the other way around does not show promising performance. These figures are in concordance with the visual demonstration of the normal versus abnormal region selection and the pathologist-approved ground-truths shown for two sample cases in Figures  \ref{fig:5_visual_example_skin} and \ref{fig:7_visual_example_breast}. These findings prove that patch-level classification performance is an indicator of how a normal atlas would perform in an indexing and search pipeline. This calls for the need to validate the performance of the normal atlas at the patch level before applying it in the pipeline.

There are some limitations to this study. As previously stated, the indexing and search performance for some cases was suboptimal before applying the normal atlas. Examples of such cases include moderately versus differentiated cases of cSCC and lobular versus ductal carcinoma of the breast. We speculate that this is an active research area involving WSI representation. Improving the performance of WSI search engines and pathology-specific deep networks is not in the scope of this study. We selected the currently established pipeline of Yottixel as the state-of-the-art search engine for our experiments.

Looking toward future research, other studies have shown the role of vision transformers in the area of anomaly detection \cite{vit_anomaly_detection, vit_anomaly_detection_2, vit_anomaly_detection_3}. These approaches mainly rely on using the reconstruction score of a previously trained vision transformer encoder and decoder to extinguish normal versus abnormal image features. This approach hypothesizes that a coupled encoder and decoder trained only on images of normal phenomena would struggle to reconstruct an image of an abnormal instance. This approach holds promise for training unsupervised anomaly detection models in medical imaging and should be further investigated in future studies.
\bibliography{bibliography}
\section*{Acknowledgements}
Some figures are created with BioRender.com.
\clearpage
\onecolumn
\appendix
\section*{Supplementary Materials}
\begin{table*}[ht!]
\centering
\caption{Performance metrics for different methods skin}
\label{tab:metrics_table_skin}
\begin{tblr}{
  width = \linewidth,
  colspec = {Q[154]Q[206]Q[69]Q[56]Q[54]Q[54]Q[54]Q[265]},
  cells = {c},
  cell{2}{1} = {r=9}{},
  cell{2}{2} = {r=3}{},
  cell{5}{2} = {r=3}{},
  cell{8}{2} = {r=3}{},
  cell{11}{1} = {r=9}{},
  cell{11}{2} = {r=3}{},
  cell{14}{2} = {r=3}{},
  cell{17}{2} = {r=3}{},
  cell{20}{1} = {r=9}{},
  cell{20}{2} = {r=3}{},
  cell{23}{2} = {r=3}{},
  cell{26}{2} = {r=3}{},
  hline{1-2,11,20,29} = {-}{},
  hline{5,8,14,17,23,26} = {2-8}{},
}
\textbf{Deep Network} & \textbf{Experiment Setup} & \textbf{Top N} & \textbf{WD} & \textbf{MD} & \textbf{PD} & \textbf{NL} & \textbf{Weighted Average F1 Score}\\
KimiaNet & No Normal Atlas & Top 1 & 0.86 & 0.35 & 0.58 & 0.96 & 0.77\\
 &  & Top 3 & 0.88 & 0.35 & 0.67 & 0.96 & 0.79\\
 &  & Top 5 & 0.89 & 0.32 & 0.67 & 0.97 & 0.80\\
 & {Normal Atlas\\Using Isolation Forest} & Top 1 & 0.84 & 0.34 & 0.59 & 0.93 & 0.75\\
 &  & Top 3 & 0.88 & 0.38 & 0.62 & 0.92 & 0.78\\
 &  & Top 5 & 0.89 & 0.42 & 0.70 & 0.94 & 0.81\\
 & {Normal Atlas\\Using One-Class SVM} & Top 1 & 0.81 & 0.38 & 0.59 & 0.76 & 0.71\\
 &  & Top 3 & 0.85 & 0.47 & 0.66 & 0.79 & 0.77\\
 &  & Top 5 & 0.85 & 0.45 & 0.69 & 0.79 & 0.76\\
ViT DINO & No Normal Atlas & Top 1 & 0.82 & 0.31 & 0.58 & 0.97 & 0.74\\
 &  & Top 3 & 0.86 & 0.35 & 0.66 & 0.99 & 0.78\\
 &  & Top 5 & 0.88 & 0.33 & 0.71 & 0.98 & 0.80\\
 & {Normal Atlas\\Using Isolation Forest} & Top 1 & 0.80 & 0.25 & 0.44 & 0.78 & 0.67\\
 &  & Top 3 & 0.83 & 0.25 & 0.41 & 0.79 & 0.69\\
 &  & Top 5 & 0.85 & 0.22 & 0.41 & 0.82 & 0.71\\
 & {Normal Atlas\\Using One-Class SVM} & Top 1 & 0.81 & 0.26 & 0.53 & 0.84 & 0.70\\
 &  & Top 3 & 0.84 & 0.27 & 0.61 & 0.86 & 0.73\\
 &  & Top 5 & 0.85 & 0.30 & 0.62 & 0.88 & 0.75\\
ResNet50 DINO & No Normal Atlas & Top 1 & 0.83 & 0.30 & 0.48 & 0.95 & 0.73\\
 &  & Top 3 & 0.84 & 0.18 & 0.50 & 0.95 & 0.72\\
 &  & Top 5 & 0.87 & 0.20 & 0.61 & 0.95 & 0.75\\
 & {Normal Atlas\\Using Isolation Forest} & Top 1 & 0.50 & 0.26 & 0.37 & 0.41 & 0.43\\
 &  & Top 3 & 0.54 & 0.22 & 0.42 & 0.41 & 0.46\\
 &  & Top 5 & 0.54 & 0.18 & 0.48 & 0.41 & 0.46\\
 & {Normal Atlas\\Using One-Class SVM} & Top 1 & 0.78 & 0.16 & 0.38 & 0.82 & 0.65\\
 &  & Top 3 & 0.83 & 0.17 & 0.36 & 0.83 & 0.68\\
 &  & Top 5 & 0.84 & 0.16 & 0.36 & 0.82 & 0.69
\end{tblr}
\end{table*}

\begin{table*}[ht!]
\centering
\caption{Performance metrics for different methods Breast}
\label{tab:metrics_table_breast}
\begin{tblr}{
  width = \linewidth,
  colspec = {Q[165]Q[223]Q[73]Q[58]Q[58]Q[58]Q[287]},
  cells = {c},
  cell{2}{1} = {r=9}{},
  cell{2}{2} = {r=3}{},
  cell{5}{2} = {r=3}{},
  cell{8}{2} = {r=3}{},
  cell{11}{1} = {r=9}{},
  cell{11}{2} = {r=3}{},
  cell{14}{2} = {r=3}{},
  cell{17}{2} = {r=3}{},
  cell{20}{1} = {r=9}{},
  cell{20}{2} = {r=3}{},
  cell{23}{2} = {r=3}{},
  cell{26}{2} = {r=3}{},
  hline{1-2,11,20,29} = {-}{},
  hline{5,8,14,17,23,26} = {2-7}{},
}
\textbf{Deep Network} & \textbf{Experiment Setup} & \textbf{Top N} & \textbf{DC} & \textbf{LC} & \textbf{NL} & \textbf{Weighted Average F1 Score}\\
KimiaNet & No Normal Atlas & Top 1 & 0.88 & 0.30 & 1.00 & 0.79\\
 &  & Top 3 & 0.89 & 0.19 & 0.98 & 0.77\\
 &  & Top 5 & 0.90 & 0.13 & 0.98 & 0.77\\
 & {Normal Atlas\\Using Isolation Forest} & Top 1 & 0.88 & 0.34 & 0.89 & 0.79\\
 &  & Top 3 & 0.89 & 0.21 & 0.86 & 0.77\\
 &  & Top 5 & 0.89 & 0.07 & 0.83 & 0.74\\
 & {Normal Atlas\\Using One-Class SVM} & Top 1 & 0.89 & 0.32 & 0.93 & 0.79\\
 &  & Top 3 & 0.89 & 0.18 & 0.81 & 0.76\\
 &  & Top 5 & 0.89 & 0.13 & 0.81 & 0.76\\
ViT DINO & No Normal Atlas & Top 1 & 0.88 & 0.39 & 1.00 & 0.80\\
 &  & Top 3 & 0.89 & 0.24 & 1.00 & 0.78\\
 &  & Top 5 & 0.90 & 0.17 & 1.00 & 0.78\\
 & {Normal Atlas\\Using Isolation Forest} & Top 1 & 0.88 & 0.37 & 0.73 & 0.78\\
 &  & Top 3 & 0.89 & 0.29 & 0.69 & 0.77\\
 &  & Top 5 & 0.89 & 0.09 & 0.69 & 0.74\\
 & {Normal Atlas\\Using One-Class SVM} & Top 1 & 0.87 & 0.37 & 0.55 & 0.77\\
 &  & Top 3 & 0.88 & 0.29 & 0.44 & 0.76\\
 &  & Top 5 & 0.88 & 0.21 & 0.25 & 0.74\\
ResNet50 DINO & No Normal Atlas & Top 1 & 0.87 & 0.27 & 1.00 & 0.77\\
 &  & Top 3 & 0.88 & 0.15 & 1.00 & 0.76\\
 &  & Top 5 & 0.89 & 0.11 & 1.00 & 0.76\\
 & {Normal Atlas\\Using Isolation Forest} & Top 1 & 0.86 & 0.17 & 0.73 & 0.74\\
 &  & Top 3 & 0.88 & 0.15 & 0.60 & 0.74\\
 &  & Top 5 & 0.88 & 0.11 & 0.32 & 0.72\\
 & {Normal Atlas\\Using One-Class SVM} & Top 1 & 0.88 & 0.31 & 0.89 & 0.78\\
 &  & Top 3 & 0.88 & 0.20 & 0.89 & 0.77\\
 &  & Top 5 & 0.89 & 0.16 & 0.76 & 0.76
\end{tblr}
\end{table*}

\clearpage
\end{document}